# A semantic cache for enhancing Web services communities activities: Health care case Study

Hela Limam
Department of Computer Science
High Institute of Management
Tunis, Tunisia

Jalel Akaichi
Department of Computer Science
High Institute of Management
Tunis, Tunisia

*Abstract*—Collective memories are strong support for enhancing the activities of capitalization, management and dissemination inside a Web services community. To take advantages of collective memory, we propose an approach for indexing a health care Web services community's' resources with semantic annotations explaining and formalizing its informative content. Then we show how the health care Web services community' members exploit their collective memory by expressing their queries allowing them searching relevant resources in order to perform their activities.

## I. INTRODUCTION (HEADING 1)

Despite their visible advantage and accessibility, the rapid growing number of published Web services prevents users or requestors from finding easily and efficiently the services relevant to their specific needs. Hence, the concept of communities of Web services has emerged for gathering Web services according to their functionalities in order to ease and improve the process of Web services discovery in an open environment like the Internet. By providing a centralized access to several functionally-equivalent Web services via a unique endpoint, communities enable processing complex users' queries that a single Web service can not satisfy.

The cornerstone of building Web services communities is their ability to be queried transparently and easily by users, which aim to satisfy their informational needs in a satisfactory time and in a pertinent retrieval. Nevertheless, processing a user query is not an easy task and may involve the access to a number of distributed communities in order to locate Web services that are capable of answering the query. Those queries are sometimes complex and short-living. Hence, it seems to be beneficial to conserve them for a future reuse and in order to be shared by communities' members who have similar informational needs.

In this context, collective memories appear to be very attractive to use in order to enhance sharing useful dedicated reusable fragments of know-how inside a Web services community. Enhancing Web services communities activities using a semantic cache memory highlights the interest of capitalizing formulated queries to the cache memory and in general the expert how-know of the community in the field of the information discovery. Hence, the process of researching resources for answering queries becomes based on a formal manipulation of annotated resources.

In our paper we propose a model for Health care services communities which enables semantic caching of queries, we provide a formal description of queries and the cache content and we detail query processing inside a community using the semantic cache.

The rest of this paper is organized as follows: In Section 2, we review previous research on semantic caching as well as other related issues. A Health care community model is defined in Section 3. Section 4 proposes a semantic cache model suitable for Web services communities. Section 5 discusses the semantic cache organization and the semantic caching query processing strategies. Finally, we summarize our work and discuss future research in Section 6.

## II. RELATED WORKS

A review of research projects aiming at assisting community activities by a collective memory as the european project SevenPro [1], the projects ANR e-WOK HUB [2] and Immunosearch [3] or the projet C3R [4] highlights the need to capitalize requests made by users in suitable databases to allow their authors to reuse or exchange them with other community members. More generally, capitalization approaches to information retrieval becomes a real issue in many areas.

Indeed, specific strategies are implemented by experts in a specific field in order retrieve information necessary for their activities [5] and are often difficult to acquire by novice users. These strategies, more and more critical with the increasing specialization and knowledge bases are capitalized and are rarely used in either the search tools like Google or in the portal domain [5].authors in [5] propose an approach to clarify the critical procedures of information retrieval in the medical field using what they call the search strategies portals .Starting with a set of standard questions in the field, they define a set of patterns representing research procedures. A search procedure is represented by an ordered set of subgoals and for each search procedure, links to relevant sources of information are established

Authors [6] offer a browsing environment of Web resources. They distinguish among three levels of knowledge: (i) a medium level of knowledge that brings web resources in the field of application (ii) Represents a level of knowledge that brings together the meta-data resources of the previous level and (iii) a level of knowledge transmission that offers courses (called "e-course") for web resources through the meta-data associated with them. E-courses are composed of steps characterized by an intention or a title, subject and an illustration. The illustrations are Web resources related to e-course step.

Examples of research procedures for dynamic navigation systems also exist in online learning-based models of semantic Web. In [7], a model for the pedagogical approach is adopted by an assembly of requests parameterized and resources whose annotations meet these requests up educational material are presented dynamically to the learner navigating through the system.

In conclusion we can say that there has been much interest in the area of applying semantic caching in Web services and communities in general. Some of the proposed approaches only work in the field Web communities while others limit queries to Web services. Hence previous works either ignore the possibility of applying semantic cache for enhancing and performing query processing among communities of Web services .The objective of our work is to extend the existing semantic caching work along several dimensions. First we present a formal semantic caching model for Web services communities, and then we explore the semantic caching query processing strategies. We examine how to efficiently answer queries against the cache. Finally, we validate semantic cache performance through a detailed Health care case study.

### III. HELTH CARE WEB SERVICES COMMUNITIES MODEL

Communities of Web services are virtual spaces that can dynamically gather different Web services having complementary functionalities in order to provide composite services with high quality. Some approaches have been proposed to organize communities of Web services. In a previous work we proposed a Web services communities design language, called WSC-UML[12], which increases the expressiveness of UML for Web services communities and guides their design. Stereotypes and graphical annotations have been added to UML diagrams in order to distinguish between the different aspects in a Web services community. WSC-UML was used to model Web services communities in general. The studied formalism is suitable for modeling a Health care community in a way that enables querying Health care Web services.

In this section we introduce a WSC-UML model for a Health care community and we specify its associated Web services. The Health care community is designed to combine data from the large ancillary services, such as pharmacy, laboratory, and radiology, with various clinical care Services .It has an identifier and is described by a set of attributes. It is composed of a set of Health care Web services: Insurance Service, Care Service, Patient Refferal Service, Physician Refferal Service and scheduling Service. Web services insider a community are associated to each other's with peer relationships. Each Web service modeled as a class in WSC-UML class diagram and is described by a set of attributes as shown in figure 1. The number of integrated Web services involved in the Health care community is dependent upon the data structures and has to provide an interface that allows clinicians to access the silo systems through a portal.

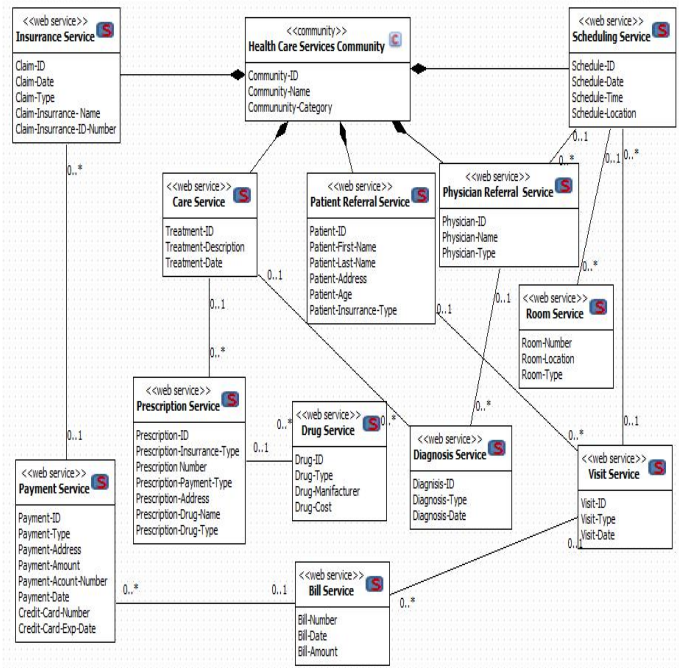

Figure 1. WSC-UML Class Diagram of a Health care community

### IV. A SEMANTIC CACHE MODEL FOR WEB SERVICES COMMUNITIES

#### A. The system general architecture

We propose a model for building, reusing and sharing queries for a pertinent information retrieval. The proposed model takes as input queries expressed by users then transforms them into a format that allows their reuse. We focus on the construction the community semantic cache that can be seen as episodic memories in which the research approaches are built dynamically based on queries. We also focus on the scenario of the information retrieval that we call the process of finding information. The general architecture of our system is exposed in figure 2.

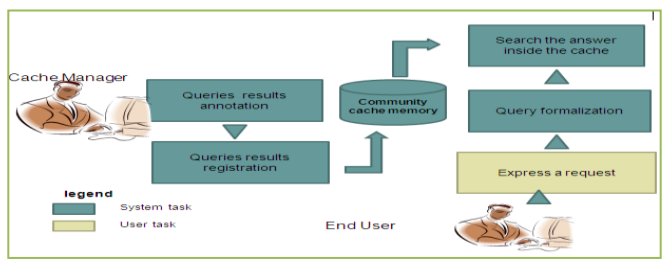

Figure 2. The semantic cache general architecture

#### B. Formal definition of the semantic cache content

The queries formulation in our work is partially inspired by the work [8]. Our approach distinguishes itself by the fact that it extends and adapts the cited work in the field of Web service communities. We consider that queries addressed to communities take the form of select and project queries, although the proposed model can be extended to handle other more complicated queries .the formal definition of the semantic cache is given in this section, later we discuss how to store and organize the semantic cache. The Web services community model presented in the previous section is considered a relational database where each modeled class is a relation. For example the class Web service is a relation. Each relation consists of a relation schema and a relation instance.

An instance of the relation << Community Member>> is a set of members satisfying the constraint of having the same number of the attributes described in the Community Member schema.

Suppose that the examined community C consists of a set of modeled classes cl1, cl2,..., cln

$$C=\{ cl_i , 1\leq i \leq n \}$$

Further let Acli stand for the attribute set defined by the schema of the class cli and A present the attribute set of the whole community, then we have:

$$A=\cup A_{cli}, 1\leq i \leq n$$

Before defining a semantic cache, we first give the predicates, through which a semantic cache is constructed.

**Definition 1:** Given a community $C=\{cl_i\}$ and its attributes set $A =\cup A_{cli}, 1\leq i \leq n$, *a* **Compare Predicate** *of* C, P, is of the form P=a op c,
where a $\epsilon$ A, op $\epsilon$ $\{\leq, <, \geq, >, =\}$,c is domain value or a constant.

In fact, a semantic cache is used to store annotated results of queries. It is composed of a set of Semantic Query Result. A Semantic Query Result is an original, decomposed, or coalesced query result. Its definition is consistent with that of a materialized view [9]. To further simplify the problem, we assume that the selection condition of a query is an arbitrary constraint formula of compare predicates, namely, a disjunction of conjunctions of compare predicates.

**Definition 2:** Given a community $C= \{cl_i\}$ and its attributes set $A=\cup A_{cli}, 1\leq i \leq n$, a Semantic Query Result is a tuple $<C_C, S_A, S_P, S_C>$ where $S_C=\pi_{S_A}\sigma_{SP}(C_C)$, $S_R \epsilon C$, $S_A \subseteq A_{CC}$ and $S_P = P_1 \vee P_2 \vee \ldots P_m$ where each $P_j$ is a conjunctive of comparative predicates, i.e., $P_j=b_{j1} \wedge b_{j2} \wedge b_{jl,}$
Each $b_{jl}$ is a compare predicate involving only the attributes in $A_{CC}$.

In definition 2, CC and SA define the class and attributes involved in computing S, respectively, SP indicates the select condition that the tuples in S satisfy. Hence, these three elements specify the semantic information associated with S. The actual content of S is represented by SS. From the restrictions added, we can see that semantic query results are the results of Select-Project operations, with the selection conditions containing only compare predicates.Before queries get answered, their contents are empty (i.e., QC=Φ). Therefore, we formally define a query just as we define a semantic query result.

**Definition 3:** A Query Q has the form $Q= <Q_C,Q_A,Q_P ,Q_S >$
A semantic cache is defined as a set of semantic queries results. To reduce space overhead, the cached semantic queries results do not overlap with each other. In the following, we first give the concept of disjointed queries results, and then formally define a semantic cache.

**Definition 4:** Two Semantic queries results $S_i =<S_{iC} ,S_{iA} ,S_{iP} ,S_{iS} >$ and $Sj =<S_{jC} ,S_{jA} ,S_{jP} ,S_{jS} >$ are said to be disjointed if and only if :

1. $S_{iA} \cap S_{jA} =\Phi$ or

2. $S_{iP} \wedge S_{jP}$ is unsatisfiable.

**Definition 5:** A Semantic Cache, S C,is defined as SC={semantic query result $S_i$} where $\forall$ j, k($S_j \epsilon SC \wedge S_k \epsilon SC \wedge j \neq k$ ) Sj and Sk are disjointed).

*C. Semantic cache organization*

Several approaches tackle the problem of the physical storage of semantic queries results. The work in [10]stores queries results in tuples, and associates every query with a pointer to a linked list of the corresponding tuples. This approach works fine for select-only queries and memory caching. The key advantage is easy maintenance: tuples can be added, deleted, or moved between segments conveniently. However, this linked list scheme is not appropriate for disk caching, since it may result in too many I/O operations.

Moreover, when select-project queries are cached, the resulting tuples for different segments are no longer at the same length. Hence, even for memory caching, its advantage in maintenance is lost. Another noticeable disadvantage for this approach is the large space overhead caused by the tuple pointers.

In our case, semantic cache is composed of two parts: the content and the index. Every semantic query result is stored in one or multiple linked pages, and is associated with a pointer pointing to its first page in the memory (disk) cache. Each page contains a query result, rather than the community classes. The cache space is also managed at a page level, which makes semantic query result allocation and deallocation algorithms more straightforward and simpler. For allocation, if there are enough free pages to hold a query result, and then allocate the pages to it; for deallocation, just mark the deallocated pages as free. The index part maintains the semantic as well as physical storage information for every cached query result. In what follows, we list the basic items kept in the index. For every cached query result, we have:

- The name S, the community class $C_C$, the attribute set $S_A$, and the selection predicate $S_P$
- The pointer pointing to the first page that stores the query result
- The timestamp indicating when the query result was last visited $S_{TS}$

The index structure proposed here is consistent with the formal definition of the semantic cache. In addition to the four basic components of the semantic query result, we further add other items for maintenance use, such as $S_{TS}$. The semantic cache index is more clearly illustrated through the following Example 1.

**Example 1:** Consider a health care community with two classes: Patient referral service and scheduling services

Patient referral service (Patient-ID, Paddress, Ptelephone, Pfirst-name, Plast-name, P Age, Pinsurance-Type) and

Scheduling Services (Schedule-ID, Sdate, Stime, Slocation); also suppose that the cache contains four queries result:

- $S_1$: Select Plast-name From Patient referral service Where 20 < PAge < 60;
- $S_2$: Select Pfirst-name, Plast-name From Patient referral service Where PAge> 10;
- $S_3$: Select Schedule-ID From Scheduling Services Where SDate= 28/12/2010;
- $S_4$: Select Slocation From Scheduling Services Where Sdate > 10/12/2010 ;

Also, suppose that the first pages of the four queries results are one, three, five, and six, respectively, and $S_i$ was last visited at $T_i$, then the index is shown in Table 1.

TABLE 1. QUERIES RESULTS ORGANIZATION IN SEMANTIC CACHE

| S | $C_C$ | $S_A$ | $S_P$ | $S_C$ | $S_{TS}$ |
|---|---|---|---|---|---|
| S1 | Patient referral service | Plast-name | 20 < PAge < 60 | 1 | $T_1$ |
| S2 | Patient referral service | Pfirst-name, Plast-name | PAge> 10 | 3 | $T_2$ |
| S3 | Scheduling Service | Schedule-ID | SDate= 28/12/2010 | 5 | $T_3$ |
| S4 | Scheduling Service | Slocation | Sdate > 10/12/2010 | 6 | $T_4$ |

## V. SEMANTIC CACHING AND QUERY PROCESSING

To process a query from a semantic cache, we first check whether it can be answered by the cache. If yes, the locally available results are computed directly from the cache. When the query can only be partially answered, we trim the original query by removing or annotating the already answered parts and send it to the database server for processing. In this section, algorithms for semantic caching query processing are examined.

### A. Theoretic Foundation

From the concept of Derivability defined in [11] we introduce the following definition.

**Definition 6:** Consider a semantic query result $S=<C_C, S_A, S_P, S_C>$ and a query $Q=<Q_C, Q_A, Q_P, Q_S>$, we say Q is answerable from S, if there exist a relational algebra expression F containing only project and select operations, and only involving attributes in $S_A$, such that $F(S_C) \neq \Phi$ and $\forall$ t ($t \in F(S_C)$) $\Rightarrow$ ( t satisfies $Q_P \wedge$ t contains only attributes in $Q_A$)). Furthermore, if $F(S_C)=Q_C$, we say Q is fully answered from S; otherwise, we say Q is partially answered from S.

From Definition 6, we know that the key to compute a query from a cached segment is to find the function F, and to make sure that F can be executed on the segment. Sometimes, even the entire result of a query Q is contained in a segment S, Q still is not answerable from S.

This is because some of the attributes needed in F cannot be found in S. So, in Definition 6, we add an additional restriction on F. The following Example 2 illustrates such a point.

**Example 2:** Consider health care community and the semantic cache described in Example 1, suppose there comes a query Q = Select Pfirst-name, Plast-name From Patient referral service Where( PAge> 5) $\wedge$ (Patient-insurance-type= Personal). Obviously, the result of Q is totally contained in $S_2$, since every tuple which satisfies (PAge > 10) $\wedge$ (Patient-insurance-type= Personal) will always satisfy (PAge > 5). However, Q cannot be computed from $S_2$, because we cannot find an F as specified in Definition 6.

Intuitively, Q seems to be computed from $S_2$ by a function π Pfirst-nameσ (PAge > 10) $\wedge$ (Patient-insurrance-type= Personal)

But the attribute "Patient-insurrance-type" used in this function is not in $S_2$ after the projection.

**Definition 7:** Consider a query $Q= <Q_C, Q_A, Q_P, Q_S>$, the predicate attribute set, $Q_{PA}$, contains all the attributes that occur in $Q_P$, i.e., $Q_{PA} = \{$ a $|$a is an attribute, and a occurs in $Q_P\}$. Consider a semantic query result $S =<C_C, S_A, S_P, S_C>$, a query $Q= <Q_C, Q_A, Q_P, Q_C>$, and Q's predicate attribute set $Q_{PA}$. Then we have:

- Statement 1: If $C_C = Q_C$, $S_A \cap Q_A \neq \Phi$, $Q_P \wedge S_P$ is satisfiable by $C_C$, and $Q_{PA} \subseteq S_A$, then Q is answerable from S.

- Statement 2: If $C_C = Q_C$, $Q_A \subseteq S_A$, $Q_P \Rightarrow S_P$, and and $Q_{PA} \subseteq S_A$, then, Q is fully answered from S.

**Definition 8:** Consider a semantic segment $S =<C_C, S_A, S_P, S_C>$. Let Y be the set of all attributes uniquely determined by the attributes in the attribute set X, with respect to SP. If X $\subseteq SA$, we say S is an extensible semantic segment, $S_A \cup Y$, denoted by $S_A^+$ is called the extended attribute set of S, and the semantic query result $S= <C_C, S_A, S_P, S_C>$ is called the extended query result of S. Since $S_A$ is uniquely determined by X with respect to $S_P$, if a tuple consisting of attributes in X satisfies $S_P$, when extended to contain attributes in $S_A^+$, it will also satisfy $S_P$.

This makes it possible to extend S to $S^+$ Notice that for each extensible query result, there could exist multiple different extended attribute sets, and hence multiple different extended squeries results. To investigate how to use extended segments in query processing, we examine Example 2 again. Suppose "Pfirst-name" is the key of "Patient referral service" class, thus other attributes of "Patient referral service" can be uniquely determined by "Pfirst-name" Hence, $S_2$ is an extensible query result. Clearly, "Patient-insurance-type" can be uniquely determined by "Pfirst-name" To form $S_2^+$, we retrieve the tuples containing both "Pfirst-name" and "Patient-insurance-type" from the community knowledge base append them to $S_{2C}$ according to the value of "Pfirst-name". After that, Q can be computed from $S_2^+$. Therefore, we have:

**Statement 3**: Consider an extensible query result $<C_C,S_A,S_P,S_C>$ and a query $<Q_C,Q_A,Q_P,Q_S>$ and suppose $SA+$ is an extended attribute set of S, and $S+$ is the extended query result of S with respect to $SA+$, QPA is Q's predicate attribute set. Then, we have:

1. If $C_C= Q_C$, $S_A^+ \cap Q_A \pm \Phi$, $Q_P \wedge S_P$ is satisfiable, and $Q_{PA} \subseteq S_A^+$, then Q is answerable by $S^+$.

2. If $C_C= Q_C$, $Q_A \subseteq S_A$, $Q_P \Rightarrow S_P$, and $Q_{PA} \subseteq S_A^+$, then Q can be fully answered by $S^+$.

**Definition 9:** Consider a semantic segment $S=<C_C,S_A,S_P,S_C>$, suppose $K_A$ is the primary key of $C_C$. If $K_A \subseteq SA$, we say S is a key-contained query result.

**Definition 10:** Consider a key-contained query result $S=<C_C,S_A,S_P,S_C>$, and a query $=<Q_C,Q_A,Q_P,Q_S>$ and suppose $Q_{PA}$ is its predicate attribute set. Then, we have:

1. If $C_C= Q_C$, $Q_P \wedge S_P$ is satisfiable, then Q is answerable by $S+< C_C, S_A,Q_A \quad Q_{PA},S_P,S^+_C >$.

2. If $C_C= Q_C$, $Q_P, S_P$, then Q can be fully answered by $S^+ = < C_C, S_A,Q_A,Q_{PA},S_P,S^+_C >$.

*B. Query processing*

Since a community does not store Web services locally, processing the query requires locating Web services that are capable of answering the query. These Web services can be selected from the local members of the community or from the semantic cache. We propose a collaborative query processing technique that consists of two steps:

- Dividing the query into parts when put together, satisfy all constraints expressed in the query
- Resolving the query by sending it to the selected parts.

For the first step, we adopt a query rewriting algorithm, which takes as input the community classes $S=<C_C,S_A,S_P,S_C>$

>, and the query $Q =<Q_C,Q_A,Q_P,Q_S>$ then produces the following output:

- Qlocal: the part of the query Q that can be answered by the community's local semantic queries results S, that is, the attributes specified in the query that are supported by the local members. It also gives the combination of the local members that can answer all (or part of) the query.

- Qrest: the part of the query that cannot be answered by the local queries results. The community will identify any external members who can answer this part of the query. Hence, Qrest is forwarded to peers. The expected answers of the forwarding is the combination of the external members that are capable of answering Qrest.

Relationship between Q and S fall into five types as described in figure 3.

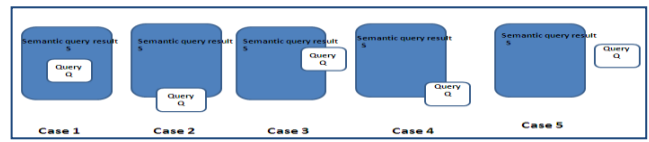

Figure 3. Query and semantic query result relationships

To summarize the query rewritwing work discussed in this section, we present Query Rewriting Algorithm, which rewrites a query Q via a key-contained semantic query result S which is defined on the same relation as Q.

**Query Rewriting Algorithm**

```
QueryRewrit (Query Q, Seamtic query result
S, Query lq, Query aq, Query rq1, Query
rq2, int type), to rewrite a query Q via a
semantic query result S.
Input: Query Q; key-contained semantic
query result S
Output: local Query lq; Amending Query aq;
Rest Query rq1, rq2; Tr Type type
    Procedure: {
    KA←S's key attribute set;
    A1←(QA∩SA) ∪KA;A2←(QA-SA)∪KA;
    IF QA⊆SA {
    IF  QP⇒SP  {
    /***** Case 1*****/
    type = 1;
    IF (QPA⊆SA) THEN aq = NULL; ELSE
    aq = πKA σQP (QR)†;
    lq = πQA KA σQP (SC)†
    rq1 = rq2 = NULL; return; }
    IF (QP∧ SP is satisfiable) {
    /***** Case 2*****/
    type = 2;
    IF (QPA⊆SA) THEN aq = NULL; ELSE
    aq = πKAσ QP∧SP (QR)†;
```

```
lq = πQA∪KA σ QP (SC)†;
rq1 = πQA∪KA σ QP∧¬SP(QR)†;
rq2 = NULL; return; } }
IF(QA⊆SA)does not hold {
IF (QP⇒SP )† {
/***** Case 3 *****/
type = 3; lq =π A1(SC)□; rq1 =π A2 σQP
(QR)†;
rq2 = aq = NULL; return; }

IF (QP∧SP is satisfiable) {
/***** Case 4 *****/
type = 4;
lq = πA1(SC);rq1=πQAπKA σ QP∧¬SP(QR)
rq2 = πA2σQP∧SP (QR)†; aq = NULL;
return; } }
/***** Case 5 *****/
rq1 = Q; \ pq = aq = rq2 = NULL; type =
5; return; }
```

## VI. CONCLUSION

We have presented a semantic cache mechanism designed for enhancing querying a Health care community. Semantic caching is based on the semantic representation of cached data and processing queries by construction of local queries for retrieving cached data and rest queries for fetching data from remote servers. Hence ,we proposed a semantic cache architecture for caching multiple queries addressed to the community and considered all operational cases. For all types of answers we have developed algorithms for query evaluation against the cache content. In next works we tackle the problem of Web service synchronization when changes occur on Web services which may alter queries results stored in the semantic cache. We also plan to propose replacement strategies for the cache maintenance in order to tune them better to real user profiles.

[1] Cherif H,Corby O, Faron C, Khelif K," Semantic annotation of texts with RDF graph contexts" In Proceeding of the International Conference on Conceptual Structures (ICCS'2008), pp.75 -82, 2008.
[2] Khalid Belhajjame , Mathieu d'Aquin , Peter Haase , Paolo Missier, "Semantic hubs for geographical projects". In Proceeding of SemanticMetadataManagement and Applications (SeMMA),workshop at ESWC, pp. 3–17, 2008.
[3] Kefi L.,Demarkez M,Collard M, "A knowledge base approach for genomics data analysis" In Proceeding of the International Conference on Semantic Systems, Graz, Austria, 2008.
[4] Faron C,Mirbell I,Sall B, Zarli, "Une approche ontologique pour formaliser la connaissance experte dans le modèle du contrôle de conformité en construction " ,19ème journées francophones d'ingénierie des connaissances, Nancy, France,Capaudes Editions ,2008.
[5] BhavnaniS, Bichakjian C, JhonsonT, Little R, Peck F, Strecher V, "Strategy hubs: Next generation domain protals with search procedures". In Proceeding of ACM Conference on Human Factors in
[6] Buffereau B, Duchet P, Picouet P, "Generating guided tours to facilitate learning from a set of indexed resources". In Proceeding of IEEE International Conference on Advanced Learning Technologies (ICALT), pp. 492, Athens, Greece : IEEE Computer Society,2003.
[7] Yessad A, Faron C, Dieng R, LASKRI M, "Ontology-driven adaptive course generation for web-based education". In World Conference on Educational Multimedia, Hypermedia and Telecommunications (ED MEDIA), Vienna, Austria,2008.
[8] Limam Hela, Akaichi Jalel, Oueslati Wided "WSC-UML: A UML Profile for Modeling Web Services Communities: A Health Care Case Study" International Journal of Advanced Research in Computer Science, Vol.2, No.2, Mars 2011.
[9] A. Gupta , I. Singh Mumick, "Maintenance of Materialized Views: Problems, Techniques, and Applications," Data Eng. Bull.,vol. 18, no. 2, pp. 3-18, 1995.
[10] S. Dar, M.J. Franklin, B.T. Jonsson, D. Srivatava, M. Tan, "Semantic Data Caching and Replacement," In Proceeding of VLDB Conf.,pp. 330-341, 1996.
[11] P.A. Larson, H.Z. Yang, "Computing Queries from Derived Relations" In Proceeding of Very Large Databases, pp. 259-269, 1985.
[12] Limam, H., Akaichi, J., Oueslati, W.:WSC-UML: A UML Profile for Modeling Web services communities. Vol. 2 , No,2, pp; 285-290,(2011).